# Differentiating surface and bulk interactions in nanoplasmonic interferometric sensor arrays

Beibei Zeng,*[a] Yongkang Gao,[a] Filbert J. Bartoli,†[a]

Detecting specific target analytes and differentiating them from interfering background effects is a crucial but challenging task in complex multi-component solutions commonly encountered in environmental, chemical, biological, and medical sensing applications. Here we present a simple nanoplasmonic interferometric sensor platform that can differentiate the adsorption of a thin protein layer on the sensor surface (surface effects) from bulk refractive index changes (interfering background effects) at a single sensing spot, exploiting the different penetration depths of multiple propagating surface plasmon polaritons excited in the ring-hole nanostructures. A monolayer of bovine serum albumin (BSA) molecules with an effective thickness of 1.91 $nm$ is detected and differentiated from a $10^{-3}$ change in the bulk refractive index unit of the solution. The noise level of the retrieved real-time sensor output compares favorably with traditional prism-based surface plasmon resonance sensors, but is achieved using a significantly simpler collinear transmission geometry and a miniaturized sensor footprint.

## 1. Introduction

Surface plasmon resonance (SPR) biosensors have unquestionable advantages for studying biological and chemical interactions, and have become the gold standard for *real-time*, *label-free* detection of biomolecular interactions [1-9]. However, it remains a challenge for SPR sensors to detect small concentrations of target analytes in complex solutions [10]. Several approaches have been proposed to differentiate target surface biomolecular binding from background interference effects such as non-specific adsorption and bulk refractive index (RI) variations. A two-plasmon spectroscopic approach was employed to excite two SPR modes with different penetration depths at two different locations on the sensor surface [11-13]. Another approach proposed a two-channel-SPR compensation technique using two separated sensor channels, one without surface functionalization and the other covered by surface receptors with affinity for specific analytes [14]. Although these two methods have successfully retrieved and differentiated surface and background changes, identical conditions cannot be guaranteed at two different sensing spots, and an increased number of sensor channels is required. Subsequently, a third approach was proposed utilizing self-referencing dual-mode SPR sensors, in which a broadband light source illuminates at a single sensing spot and simultaneously excites two distinct SPR modes with different penetration depths (e.g. long- and short-range SPRs [10,15], or two SPR-waveguide resonances under dual polarizations [16]). Generally, all of above approaches are restricted to the traditional prism-based SPR configuration [3], which, while successful for numerous applications, suffers from their low achievable spatial resolution and throughput, as well as its complex and bulky optical geometry.

Nanoplasmonic biosensors are being actively investigated for next-generation sensing applications that require system miniaturization, simpler optical geometry, increased spatial resolution, and high-throughput multiplexing detection [17-27]. Nanoplasmonic biosensors frequently employ metallic nanoparticles or nanohole (or nanoslit) arrays to couple incident light directly into surface plasmon polaritons (SPPs) [17-26], utilizing a much simpler collinear optical configuration, and hence offering opportunities for system miniaturization and integration. These nanoplasmonic biosensor arrays permit high-throughput multiplexed detection, when using a CCD camera for the simultaneous measurement of transmitted light intensity from multiple sensing spots, and can achieve a high spatial resolution, with sensing spots size as small as a few $\mu m^2$ [19]. However, their performance is generally limited by their broad spectral line-width, low contrast and weak resonance intensity [17-25]. The reported detection limits for multiplexed detection in these nanoplasmonic biosensors are typically one to two orders of magnitude poorer than that of commercial SPR systems [17-25]. To overcome this limitation, nanoplasmonic interferometric biosensors were recently proposed, and shown to exhibit interference patterns with narrow linewidths and high contrast ratios, significantly improving the detection resolution. These nanoplasmonic interferometric biosensors utilize phase-sensitive interferences between the free-space light and propagating SPPs to achieve superior spatial resolution and detection performance [28-30]. Using circular nanoplasmonic interferometer arrays, our group has recently demonstrated a spectral detection resolution of 0.4 $pg/mm^2$, comparable to state-of-the-art commercial SPR sensors, but that was achieved using a much simpler collinear transmission geometry and a miniaturized sensor footprint [29]. In addition, a record high sensing figure-of-merit of 146 (FOM*=(d$I/I_0$)/d$n$) can be obtained for these nanoplasmonic interferometric sensors using the intensity-interrogation CCD imaging techniques. These are quite promising for scalable high-throughput multiplexing applications.

Until now, there has been a notable absence of efforts to differentiate surface biomolecular binding from interfering background effects in nanoplasmonic biosensors. This can be attributed in part to the difficulty in exciting and detecting multiple SPPs with significantly different penetration depths in a single sensing spot on metallic nanoparticles and nanohole (or nanoslit) array sensors [17-22]. In contrast, the nanoplasmonic interferometric biosensors support multiple propagating SPPs with different penetration depths on the same sensing spot [29]. In the present work, we expand the functionality of the promising circular nanoplasmonic interferometric sensing platform and demonstrate a new self-referenced nanoplasmonic biosensor. By exploiting the different penetration depths of multiple propagating SPPs excited within the ring-hole nanostructures, surface biomolecular binding can be differentiated from bulk RI changes within a single sensing spot. A monolayer of BSA molecules with an effective thickness of 1.91$nm$ and a background change of $10^{-3}$ refractive index unit (RIU) are both successfully detected and differentiated, with a low noise level that is comparable with that obtained for traditional prism-based SPR sensors [10].

## 2. Results and discussion

A schematic of the *collinear* transmission setup is shown in Fig.1 (a). A white light beam from a 100$W$ halogen lamp illuminated the sensor chip through the condenser of an Olympus IX81 inverted microscope. The field and aperture diaphragms of the condenser were both closed to obtain a nearly collimated light beam. The transmitted light was collected by a 40× objective lens (numerical aperture, NA=0.6), and then coupled to an Ocean Optics USB4000 portable spectrometer for spectral measurements. An Indel E-beam evaporation system was used to deposit a 5$nm$-thick titanium and subsequently a 300$nm$-thick Au film onto a pre-cleaned standard Fisher Scientific glass slide. FEI Scios Dual-Beam focused ion beam (FIB) milling (Ga+ ions, 30kV, 30pA) was used to fabricate arrays of ring-hole nanostructures with a center-to-center distance of 12.5$\mu m$ on the Au film, as shown in Fig. 1(b). A 12×12 ring-hole interferometer array thus has a sensor footprint of around 150×150 $\mu m^2$. The structural parameters for each ring-hole nanostructure are the same as those reported in Ref. [29] (Fig. S1 in Supplementary materials). After the FIB milling, the sensor chip was cleaned using oxygen plasma (PX-250, March Instruments) and bonded to a PDMS microfluidic channel, as Fig. 1(c) shows. The PDMS microfluidic channels were fabricated by conventional soft lithography. A SU8-50 (Microchem) master mold of the channel (50 $\mu m$ deep and 4 mm wide) was patterned on a silicon wafer by photolithography. A 10:1 ratio of PDMS (Sylgard 184, Dow corning) and curing agent were used to cast the mold, and then baked at 70 ºC for 3 hours. The PDMS channel was cut and peeled from the master, and inlet and outlet holes were punched for tubing. The multiple circular grooves scatter the normally incident broadband light into SPPs that propagate towards the central nanohole. The groove periodicity was carefully chosen so that the SPPs launched at each groove are approximately in phase in the spectral region of interest, generating strong propagating SPP waves directed to the central nanohole. The red solid curve in Fig. 1(d) is the measured transmission spectrum of the 12×12 ring-hole interferometer array in a water environment. The multiple transmission peaks and valleys result from the constructive and destructive interference between propagating SPPs and the light transmitted directly through the central nanoholes [29]. The spectral positions of these interference peaks and valleys (e.g. $\lambda_1$ to $\lambda_6$) are very sensitive to local RI changes, caused by either surface effects (e.g. biomolecular adsorption) or background RI changes (e.g. temperature or compositional variations).

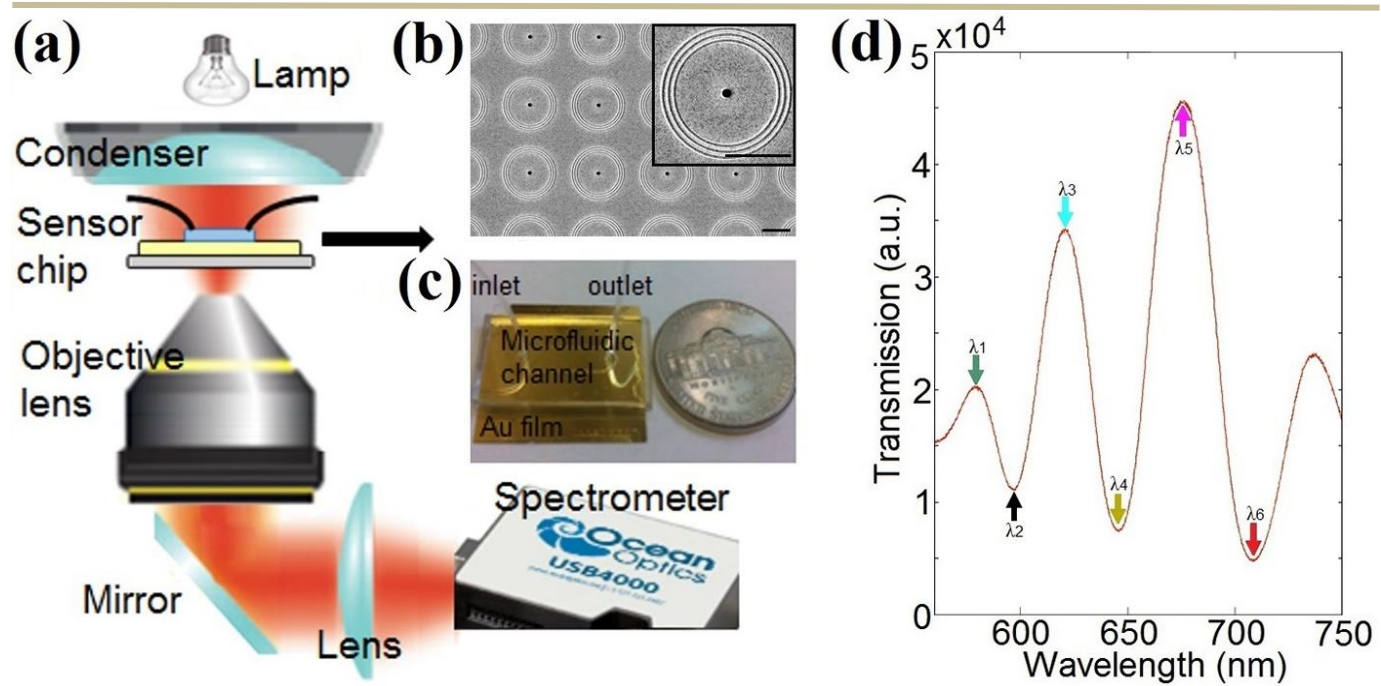

Fig.1. (a) Schematic of the optical setup. (b) Scanning electron microscope (SEM) images of the fabricated interferometric ring-hole arrays. The center-to-center distance between each ring-hole nanostructure is 12.5$\mu m$, and the sensor foot-print is 150×150$\mu m^2$. Scale bar, 5$\mu m$. The inset shows an enlarged ring-hole nanostructure. (c) A photograph of the sensor chip, consisting of 12×12 ring-hole interferometer arrays fabricated on a 300$nm$-thick Au film, integrated with a PDMS microfluidic channel and inlet/outlet tubing. (d) Measured transmission spectra for the 12×12 ring-hole interferometer arrays. The multiple transmission peaks and valleys are marked by 6 colored arrows, corresponding to wavelengths from $\lambda_1$ to $\lambda_6$.

SPPs excited at different wavelengths in the interferometric ring-hole nanostructures have different penetration depths $\delta_d$ into the surrounding aqueous medium, which are described by $\delta_d = \lambda/2\pi \cdot \sqrt{|(\text{Re}(\varepsilon_m) + \varepsilon_d)/\varepsilon_d^2|}$, where $\lambda$ is the excitation wavelength, $\varepsilon_m$ and $\varepsilon_d$ are the relative permittivity of metal and dielectric medium, respectively (Fig. S2 in Supplementary materials) [31-36]. SPPs with different penetration depths $\delta_d$ have different relative sensitivities to surface and bulk RI changes, due to differing overlap between electromagnetic fields of SPPs and the surrounding medium [10]. Generally, SPPs with shorter excitation wavelengths have smaller $\delta_d$ and are more sensitive to RI changes due to biomolecular adsorption at the sensor surface. Similarly, SPPs with larger $\delta_d$ are more sensitive to bulk RI changes [17]. Therefore, surface and bulk RI changes will generally result in different wavelength shifts for each interference peak and valley from $\lambda_1$ to $\lambda_6$. This enables the separation of surface and bulk RI changes in a single sensing spot by recording the wavelength shifts of multiple peaks or valleys. This can be accomplished simply using a single spectrometer. To describe this process, we define the surface sensitivity as the shift in wavelength caused by a 1$nm$ change in the surface layer thickness (e.g. due to biomolecular adsorption), and the bulk RI sensitivity as the shift in wavelength due to a unit change in the bulk refractive index (e.g. background solutions with different glycerol concentrations), respectively. The net sensor response, including contributions from both surface and bulk RI changes, is given by [10, 15]:

$$\Delta\lambda_m = S_m^S \cdot \Delta d_e + S_m^B \cdot \Delta n_B \quad (1)$$

where $\Delta\lambda_m$ is the wavelength shift at each interference peak or valley marked by $\lambda_m$ ($m$ ranges from 1 to 6). $S_m^S$ and $S_m^B$ are surface and bulk sensitivities for each interference peak or valley, respectively. The values for $S_m^S$ and $S_m^B$ can be determined experimentally, as described below. $\Delta d_e$ and $\Delta n_B$ represent the changes in surface layer thickness and bulk RI, respectively.

The spectral positions of each interference peak and valley ($\lambda_1$ to $\lambda_6$) were measured as a function of time as the sensor surface sequentially experienced a simple change in bulk RI, surface adsorption of BSA molecules, and combined changes in BSA surface coverage and bulk RI, as shown in Fig. 2(a). Deionized (DI) water was first injected into the PDMS microfluidic channel to rinse the sensor chip and define the baseline for the experiment. Subsequently, a 6% glycerol-water (6% G) solution was introduced into the channel, causing a bulk RI change of $\Delta n_B = 0.009\ RIU$. Since no surface adsorption occurred during this time period, $\Delta d_e = 0$. Consequently, Eq. (1) reduces to $\Delta\lambda_m = S_m^B \cdot \Delta n_B$, and the bulk sensitivity $S_m^B$ can be determined for each interference peak and valley, as shown in Fig. 2(b). Next, DI water was injected to remove the 6% G solution. A 500$\mu g/ml$ solution of BSA in water was then introduced into the channel, functionalizing the surface with a monolayer of BSA molecules. A subsequent DI water wash caused no additional wavelength shift, indicating that a saturated BSA surface monolayer had been formed with an effective thickness of $\Delta d_e = 1.9nm$ [37]. In this case, the wavelength shift $\Delta\lambda_m$ is solely due to surface binding of BSA ($\Delta n_B = 0$), and Eq. (1) becomes $\Delta\lambda_m = S_m^S \cdot \Delta d_e$. The resulting surface sensitivity $S_m^S$ was calculated for each interference peak and valley and is shown in Fig. 2(c). Finally, a 6% G solution was re-injected into the channel. The wavelength shift $\Delta\lambda_m$ for each interference peak and valley, shown in Fig. 2(d), now reflects both surface layer BSA adsorption and bulk RI changes, simulating complex solutions. By incorporating the experimentally determined surface and bulk sensitivities $S_m^S$ and $S_m^B$ into Eq. (1), we can determine the surface layer thickness and bulk RI changes, $\Delta d_e$ and $\Delta n_B$, from the measured spectral shift in the interference pattern at two different wavelengths, as we illustrate below.

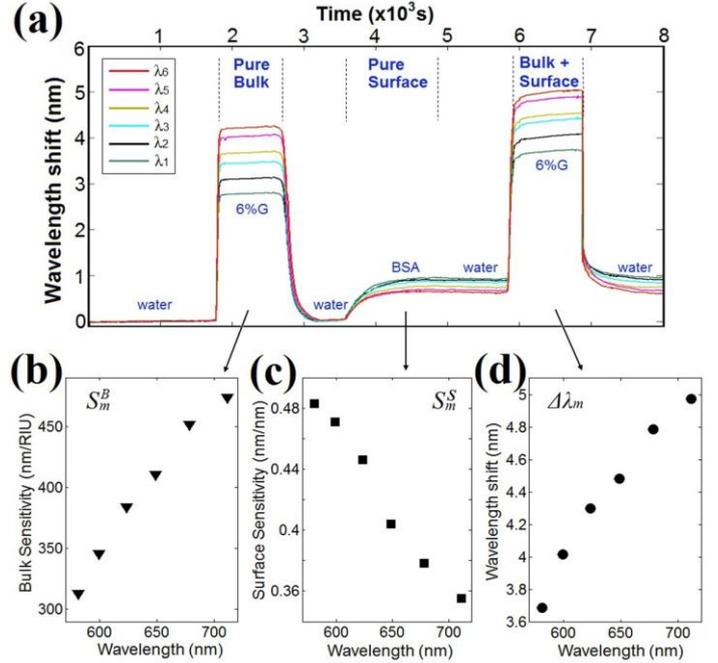

Fig.2. (a) Real-time sensor responses (wavelength shifts $\Delta\lambda_m$) of six different interference peaks and valleys (corresponding to wavelengths from $\lambda_1$ to $\lambda_6$) to bulk RI (6% G) and surface layer thickness (BSA adsorption on the sensor surface) changes. Calibrated (b) bulk $S_m^B$ and (c) surface $S_m^S$ sensitivities for each interference peak and valley from $\lambda_1$ to $\lambda_6$. (d) Wavelength shifts $\Delta\lambda_m$ of six interference peaks and valleys in complex solutions with both surface (BSA adsorption) and bulk (6% G) RI changes.

To demonstrate the capability of circular nanoplasmonic interferometric sensors to differentiate surface and bulk effects in complex media, we consider an example in which the adsorption of BSA molecules on the sensor surface is detected in the presence of bulk solutions of varying refractive index. Solutions were introduced onto the sensor chip in the following order: (1) DI water, (2) 3% G, (3) 6% G, (4) DI water, (5) 500$\mu g/ml$ BSA solution, (6) DI water, (7) 3% G, (8) 6% G, (9) DI water. The measured spectral positions of the interference minima at $\lambda_2$ and $\lambda_6$ are plotted in Fig. 3(a) as a function of time. Changes in the surface layer thickness $\Delta d_e$ and bulk RI $\Delta n_B$ are the two unknown quantities in Eq. (1), which can be directly retrieved by straightforward mathematics. For example, variations in the surface layer thickness and bulk RI causes a shift in the spectral positions of the two interference valleys initially at $\lambda_2=596.7nm$ and $\lambda_6=708.1nm$. $\Delta\lambda_2$ and $\Delta\lambda_6$ can be expressed as:

$$\Delta\lambda_2 = S_2^S \cdot \Delta d_e + S_2^B \cdot \Delta n_B \quad (2)$$
$$\Delta\lambda_6 = S_6^S \cdot \Delta d_e + S_6^B \cdot \Delta n_B \quad (3)$$

Fig. 2(b) and (c) provide the measured surface sensitivity ($S_2^S = 0.471\ nm/nm$ and $S_6^S = 0.355\ nm/nm$) and bulk sensitivity ($S_2^B = 345.8\ nm/RIU$ and $S_6^B = 469.1 nm/RIU$) for these two interference valleys. Eqs. (2) and (3) can then be solved at each point in time for the unknown changes in surface layer thickness $\Delta d_e$ and bulk RI $\Delta n_B$, yielding the following expressions:

$$\Delta d_e = (\Delta\lambda_2/S_2^B - \Delta\lambda_6/S_6^B)/(S_2^S/S_2^B - S_6^S/S_6^B) \quad (4)$$
$$\Delta n_B = (\Delta\lambda_2/S_2^S - \Delta\lambda_6/S_6^S)/(S_2^B/S_2^S - S_6^B/S_6^S) \quad (5)$$

The resulting values of $\Delta d_e$ and $\Delta n_B$ are plotted as a function of time in Fig. 3(b). The figure shows a clear differentiation between the real-time change in surface layer thickness $\Delta d_e$ (red curve) and the step-like bulk RI change $\Delta n_B$ (blue curve). The retrieved RI changes for the 3% and 6% G solutions, $\Delta n_B = 0.00437\ RIU$ and $0.00889\ RIU$, agree well with the values measured using a J. A. Woollam, V-VASE ellipsometer ($\Delta n_B = 0.004\ RIU$ and $0.009\ RIU$) [29]. The bulk RI changes in background solutions (3% and 6% G) show no influence on the retrieved change in surface layer thickness, wherein a monolayer of BSA molecules with an effective thickness of 1.91$nm$ is immobilized on the sensor surface.

The uncertainty involved in extracting changes in surface layer thickness $d_e$ and bulk RI $\Delta n_B$ from the real-time sensor output depends predominantly on uncertainty in measurement of the wavelength shift (e.g. $\Delta\lambda_2$ and $\Delta\lambda_6$), inaccuracy of the sensor calibration (e.g. $S_{2,6}^S$ and $S_{2,6}^B$), and the difference in the wavelengths, and hence penetration depths associated with the interference peaks and valleys employed in the extraction processes. The insets of Fig. 3(b) show the noise levels of the retrieved real-time values for $\Delta d_e$ and $\Delta n_B$ ($\sigma_S$=0.009$nm$ and $\sigma_B$=9.2×10$^{-6}$$RIU$, obtained using over 20 data points). These values are comparable to the noise levels ($\sigma_S$=0.009$nm$ and $\sigma_B$=6.9×10$^{-6}$$RIU$) reported for traditional prism-based SPR sensors [10]. Note that the circular nanoplasmonic interferometric sensors utilize a significantly simpler collinear transmission geometry, a miniaturized sensor footprint, and a low-cost compact spectrometer. If more closely spaced interference peaks (valleys) were used, the difference in the penetration depths $\delta_d$ of two SPP modes would be smaller, resulting in greater experimental uncertainty in differentiating bulk and surface effects [10,15]. For instance, when the adjacent interference valley at $\lambda_2$=596.7$nm$ and peak at $\lambda_3$=621.1$nm$ are selected (decreasing the difference in penetration depths), the uncertainty in the extracted values of $\Delta d_e$ and $\Delta n_B$ are approximately five times larger ($\sigma_S$=0.045$nm$ and $\sigma_B$=5.5×10$^{-5}$$RIU$). While a larger wavelength separation is preferred, the noise will increase if the chosen wavelengths are beyond the optimal spectral range for interference oscillations ($\lambda_2$<$\lambda$<$\lambda_6$). For the current geometry, the two interference valleys $\lambda_2=596.7nm$ and $\lambda_6=708.1nm$ are the optimal choice for differentiating surface and bulk effects.

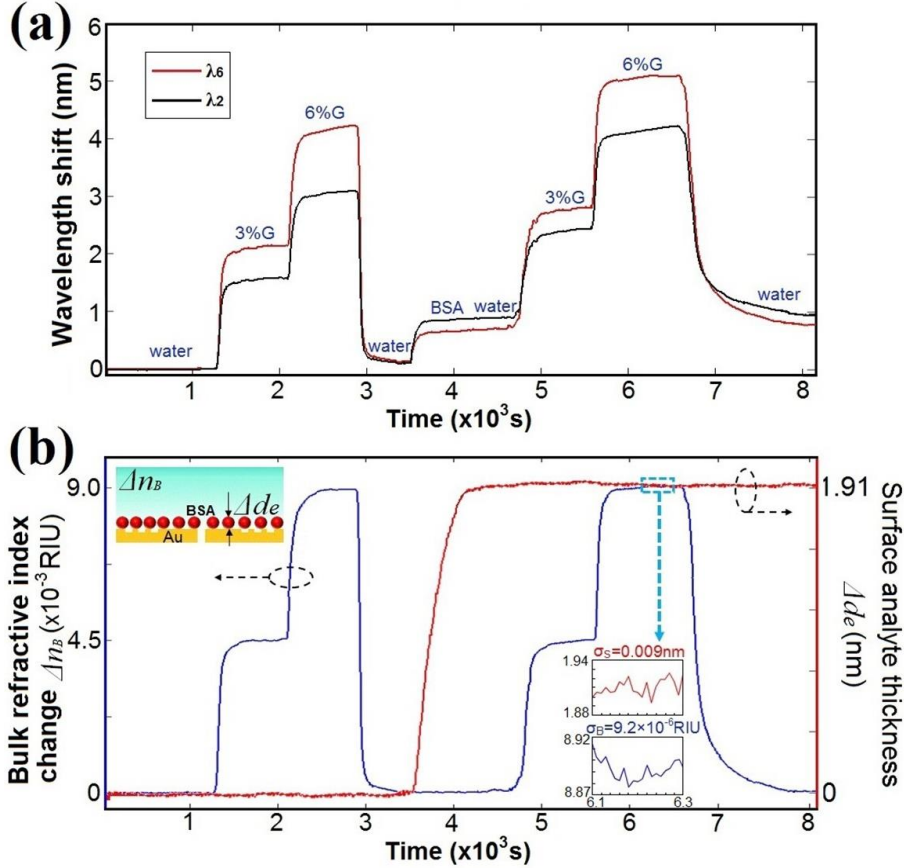

Fig.3. (a) Real-time sensor responses $\Delta\lambda_2$ (black curve) and $\Delta\lambda_6$ (red curve) to bulk RI (3% and 6% G solutions) and surface layer thickness changes (BSA adsorption) at two interference valleys of $\lambda_2$ and $\lambda_6$. (b) Surface layer thickness (red curve) and bulk RI (blue curve) changes retrieved using Eq. (4) and (5), with calculated bulk ($S_2^B$ and $S_6^B$) and surface ($S_2^S$ and $S_6^S$) sensitivities at two interference valleys of $\lambda_2$ and $\lambda_6$. The insets indicate the noise levels $\sigma_S$ and $\sigma_B$ associated with the retrieved real-time change $\Delta d_e$ in surface layer thickness and the bulk RI change $\Delta n_B$, respectively (over 20 data points).

## 3. Conclusions

A nanoplasmonic ring-hole interferometric sensor platform was shown to differentiate surface layer adsorption from interfering bulk refractive index variations based on measurements within a single sensing spot. This technique exploits the wavelength-dependent penetration depths of SPPs excited in ring-hole nanostructures, which lead to very different dependences of surface and bulk sensitivities upon wavelength. A monolayer of BSA molecules with an effective thickness of $1.91nm$ has been detected in the presence of background refractive index changes of $10^{-3} RIU$. This sensor system utilizes a simple optical configuration, employing a collinear transmission geometry and a portable spectrometer, and permits dramatically reduced sensing volumes and higher throughput assays than traditional prism-based SPR sensors. This nanoplasmonic sensor may also be used to characterize non-specific biomolecular binding events in complex biomolecular fluidics (e.g. human serum, urine, or cell cultures, *etc*.), and distinguish them from other background noise. It has the potential for significant impact on point-of-care diagnostics and personal healthcare applications, as well as other applications in biomedical, environmental and chemical detection.

## Acknowledgements
We acknowledge the financial support from National Science Foundation (CBET-1014957).

## Notes and references

[a] Electrical and Computer Engineering Department, Lehigh University, Bethlehem, PA 18015, USA.
*E-mail: bez210@lehigh.edu
†E-mail: fjb205@lehigh.edu

Electronic Supplementary Information (ESI) available: [details of any supplementary information available should be included here]. See DOI: 10.1039/b000000x/